\documentclass[english,aps,twocolumn]{revtex4}
\usepackage{epsfig}
\usepackage{graphicx}
\usepackage{amsmath}

\makeatletter
\@ifundefined{textcolor}{}
{%
 \definecolor{BLACK}{gray}{0}
 \definecolor{WHITE}{gray}{1}
 \definecolor{RED}{rgb}{1,0,0}
 \definecolor{GREEN}{rgb}{0,1,0}
 \definecolor{BLUE}{rgb}{0,0,1}
 \definecolor{CYAN}{cmyk}{1,0,0,0}
 \definecolor{MAGENTA}{cmyk}{0,1,0,0}
 \definecolor{YELLOW}{cmyk}{0,0,1,0}
 }

\makeatother

\usepackage{babel}

\begin{document}


\title{Big Bang nucleosynthesis with a
non-Maxwellian distribution}

\author{C. A. Bertulani$^{1,2}$,  J. Fuqua$^2$, and M.S. Hussein$^3$}

\affiliation{$^1$ GSI Helmholtzzentrum f\"ur Schwerionenforschung, D-64291 Darmstadt, Germany\\
$^2$ Texas A\&M University-Commerce, Commerce, TX 75429-3011, USA\\
$^3$Instituto de Estudos Avan\c{c}ados, Universidade de S\~ao Paulo, C.P. 72.012, 05508-970, S\~ao Paulo, Brazil\\ and Instituto de F\'{\i}sica, Universidade de S\~ao Paulo, C. P. 66318, 05389-970 S\~ao Paulo, Brazil}

\begin{abstract}
The abundances of light elements based on the big bang nucleosynthesis model are calculated using the Tsallis non-extensive statistics. The impact of the variation of the non-extensive parameter $q$ from the unity value  is compared to observations and to the abundance yields from the standard big bang model. We find large differences between the reaction rates and the abundance  of light elements  calculated with the extensive and the non-extensive statistics.  We found that the observations are consistent with a   non-extensive parameter $q = 1^{ + 0.05}_ {- 0.12}$, indicating that a large   deviation from the Boltzmann-Gibbs statistics ($q = 1$) is highly unlikely. 
\end{abstract}

\maketitle

\section{Introduction}
The cosmological big bang model is in agreement with many observations relevant for our understanding of the universe. However, comparison  of calculations based on the model with observations is not straightforward because the data are subject to poorly known evolutionary effects
and systematic errors. Nonetheless, the model is believed to be the only probe of physics in the early universe during the interval from $3-20$ min, after which the temperature and density of the universe fell below that which is required for nuclear fusion and prevented elements heavier than beryllium from forming.
The model is inline with the cosmic microwave background (CMB) radiation temperature of 2.275 K \cite{Not11}, and provides  guidance to other areas of science, such as nuclear and particle physics. Big bang model calculations are also consistent with the number of light neutrino families  $N_\nu =3$. According to the numerous literature on the subject, the big bang model can accommodate values between $N_\nu =1.8 - 3.9$ (see, e.g., Ref. \cite{Oli02}). From the measurement of the $Z_0$ width by LEP experiments at CERN one knows that $N_\nu = 2.9840 \pm 0.0082$ \cite{LEP06}.  

In the big bang model nearly all neutrons end up in $^4$He, so that the relative abundance of $^4$He depends on the number of neutrino families and also on the neutron lifetime $\tau_n$. The sensitivity to the neutron lifetime affects Big Bang Nucleosynthesis (BBN) in two ways. The neutron lifetime $\tau_n$ influences the weak reaction rates because of the relation between $\tau_n$ and the weak coupling
constant. A shorter (longer) $\tau_n$ means that the reaction rates remain greater (smaller) than the Hubble expansion rate until a lower (larger) freeze-out temperature, having a strong impact on the equilibrium neutron-to-proton ratio at freeze-out.
This $n/p$-ratio is approximately given in thermal equilibrium by $n/p = \exp[-\Delta m/k_B T]\sim 1/6$, where $k_B$ is the Boltzmann constant, $T$ the temperature at weak  freeze-out, and $\Delta m$ is the neutron-proton mass difference. The other influence of $\tau_n$ is due to their decay in the interval between weak freeze-out ($t\sim 1$ s) and when nucleosynthesis starts ($t\sim 200$ s), reducing the $n/p$ ratio to $n/p\sim 1/7$. A shorter $\tau_n$ implies lower the predicted BBN helium abundance. In this work we will use the value of $\tau_n=878.5 \pm 0.7 \pm  0.3$  s, according to the most recent experiments \cite{Ser05} (a recent review on the neutron lifetime is found in Ref. \cite{FG11}).   Recently, the implications of a change in the neutron lifetime on BBN predictions have been assessed in Ref. \cite{Mat05}. 

The baryonic density of the universe deduced from the observations of the anisotropies of the CMB radiation, constrains the value of the number
of baryons per photon, $\eta$, which remains constant during the expansion of the universe. Big bang model calculations are compatible with the experimentally deduced value of from  WMAP observations, $\eta = 6.16 \pm 0.15\times 10^{-10}$ \cite{Ko11}. 

Of our interest in this work is the abundances of light elements in big bang nucleosynthesis.  
At the very early stages (first 20 min) of the universe evolution, when it was dense and hot enough
for nuclear reactions to take place, the temperature of the primordial plasma decreased from a few MeV down to about 10 keV, light nuclides as $^2$H, $^3$He, $^4$He and, to a smaller extent, $^7$Li were produced via a network of nuclear processes, resulting into abundances for these species which can be determined with several observational techniques and in different astrophysical environments. 
Apparent discrepancies for the Li abundances in metal poor stars, as measured observationally and as inferred by WMAP, have promoted a wealth of new inquiries on BBN and on stellar mixing processes destroying Li, whose results are not yet final. Further studies of light-element abundances in low metallicity stars and extragalactic H II regions, as well as better estimates from BBN models are required to tackle this issue, integrating high resolution spectroscopic studies of stellar and interstellar matter with nucleosynthesis models and nuclear physics experiments and theories \cite{Fie11}.

The Maxwell-Boltzmann distribution of the kinetic energy of the ions in a plasma is one of the basic inputs for the  calculation of nuclear reaction rates during the BBN.  The distribution is based on several assumptions inherent to the Boltzmann-Gibbs statistics: (a) the collision time is much smaller than the mean time between collisions, (b) the interaction is local, (c) the velocities of two particles at the same point are not correlated, and (d) that energy is locally conserved when using only the degrees of freedom of the colliding particles (no significant amount of energy is transferred to and from collective variables and fields). If (a) and (b) are not valid the resulting effective two-body interaction is non-local and depends on the momentum and energy of the particles. Even when the one-particle energy distribution is Maxwellian, additional assumptions about correlations between particles are necessary to deduce that the relative-velocity distribution is also Maxwellian.  Although the Boltzmann-Gibbs (BG) description of statistical mechanics is well established in a seemingly infinite number of situations, in recent years  an increasing theoretical effort has concentrated on the development of alternative approaches to statistical mechanics which includes the BG statistics as a special limit of a more general theory \cite{Ts88} (see also, \cite{Ren60}). Such theories aim to describe systems with long range interactions and with memory effects (or non-ergodic systems). A very popular alternative to the BG statistics was proposed by C. Tsallis \cite{Ts88,GT04}, herewith denoted as {\it non-extensive statistics} (for more details on this subject, see the extensive reviews \cite{GT04,TGS05,Ts09}). Statistical mechanics assumes that  energy is an ``extensive" variable, meaning that the total energy of the system is proportional to the system size; similarly the entropy is also supposed to be extensive. This might be justified due to the short-range nature of the interactions which hold matter together. But if  one deals with long-range interactions, most prominently gravity; one can then find that entropy is not extensive \cite{FL01,Lim02,TS03,TS04,CS05}. 

In classical statistics, to calculate the average values of some quantities, such as the energy of the system, the number of molecules, the volume it occupies, etc, one searches for the probability distribution which maximizes the entropy, subject to the constraint that it gives the right average values of those quantities. As mentioned above, Tsallis proposed  to replace the usual (BG) entropy with a new, non-extensive quantity, now commonly called the {\it Tsallis entropy}, and maximize that, subject to the same usual constraints. There is actually a whole infinite family of Tsallis entropies, indexed by a real-valued parameter $q$, which quantifies the degree of departure from extensivity (one gets the usual entropy back again when $q = 1$). It was shown in many circumstances that the classical results of statistical mechanics can be translated into the new theory \cite{Ts09}. The importance of these families of entropies is that, when applied to ordinary statistical mechanics, they give rise to probabilities that follow power laws instead of the exponential laws of the standard case (for details on this see \cite{Ts09}). In most cases that Tsallis formalism is adopted, e.g. Ref. \cite{Pes01}, the non-extensive parameter $q$ is taken to be constant and close to the value for which ordinary statistical mechanics is obtained ($q = 1$). Some works have also probed sizable deviations of the non-extensive parameter $q$ from the unity to explain a variety of phenomena in several areas of science \cite{GT04}. 

In the next sections, we shown that the Maxwell-Boltzmann distribution, a cornerstone of the big bang and stellar evolution nucleosynthesis, is strongly modified by the non-extensive statistics if $q$ strongly deviates from the unity. As a consequence, it also affects strongly the predictions of the BBN.  There is no ``a priori"  justification for a large deviation of $q$ from the unity value during the BBN epoch. In particular, as radiation is assumed to be in equilibrium with matter during the BBN, a modification of the Maxwell distribution of velocities would also impact the Planck distribution of photons. Recent studies on the temperature fluctuations of the cosmic microwave background (CMB) radiation have shown that  a modified Planck distribution based on Tsallis statistics adequately describes the CMB temperature fluctuations measured by WMAP with $q = 1.045 \pm 0.005$, which is close to unity but not quite \cite{Ber07}. Perhaps more importantly,  Gaussian temperature distributions based on the BG statistics, corresponding to the $q\rightarrow 1$ limit, do not properly represent the CMB temperature fluctuations \cite{Ber07}. Such fluctuations, allowing for even larger variations of $q$ might occur during the BBN epoch, also leading to a change of the exponentially decaying tail of the Maxwell velocity distribution.

Based on the successes of the big bang model, it is fair to assume that it can set strong constraints on the limits of the parameter $q$ used in  a non-extensive statistics description of the Maxwell-Boltzmann velocity distribution. In the literature, attempts to solve the lithium problem has assumed all sorts of ``new physics" \cite{Fie11}. The present work adds to the list of new attempts, although our results imply a much wider impact on BBN as expected for the solution of the lithium problem. If the Tsallis statistics appropriately describes the deviations of tails of statistical distributions, then the BNN would effectively probe such tails.  The Gamow window (see figure \ref{fmaxwell}) contains a small fraction of the total area under the velocity distribution.  Thus, only a few particles in the tail of the distributions contribute to the fusion rates.   In fact,  the possibility of a deviation of the Maxwellian distribution and implications of the modification of the Maxwellian distribution tail for nuclear burning in stars have already been explored in the past \cite{MQ05,HK08,Deg98,Cor99}.  As we show in the next sections, a strong deviation from $q=1$ is very unlikely for the BNN predictions, based on comparison with observations. Moreover, if $q$ deviates from the unity value, the lithium problem gets even worse. 

\section{Maxwellian and non-Maxwellin distributions}

Nuclear reaction rates in the BBN and in stelar evolution are strongly dependent of the particle velocity distributions. The fusion reaction rates for nuclear species 1 and 2 is given by $\langle \sigma v \rangle_{12}$, i.e., an average of the fusion cross section of $1+2$ with their relative velocity, described by a  velocity distribution. It is thus worthwhile to study the modifications of the stellar reaction rates due to the modifications introduced by the non-extensive statistics.

\subsection{Non-extensive Statistics}

Statistical systems in equilibrium are  described by the Boltzmann-Gibbs entropy,
\begin{equation}
{\cal S}_{BG}=-k_B \sum_i p_i \ln p_i ,	
\end{equation}
where $k_B$ is the Boltzmann constant, and $p_i$ is the probability of the i-th microstate. For two independent systems $A$, $B$,  the probability of the system $A+B$ being in a state $i+j$, with $i$ a microstate of $A$ and $j$ a microstate of $B$, is
\begin{equation}
p^{A+B}_{i+j} = p^A_i \cdot p^B _j.
\end{equation}
Therefore, the Boltzmann-Gibbs entropy satisfies the relation
\begin{equation}
{\cal S}_{A+B}={\cal S}_A +{\cal S}_B. 
\end{equation}
Thus, the entropy based on the Boltzmann-Gibbs statistic is an {\it extensive} quantity. 

In the non-extensive statistics \cite{Ts88}, one replaces the traditional entropy by the following one:
\begin{equation}
{\cal S}_q=k_B {1-\sum_i p^q_i \over q-1},
\end{equation}
where $q$ is a real number. For $q=1$, ${\cal S}_q= {\cal S}_{BG}$. Thus, the Tsallis statistics is a natural generalization of the Boltzmann-Gibbs entropy.

Now it follows that
\begin{equation}
{\cal S}_q(A+B)={\cal S}_q(A) +{\cal S}_q(B)+{(1-q)\over k_B}{\cal S}_q(A){\cal S}_q(B). 
\end{equation}
The variable $q$ is a measure of the {\it non-extensivity}. Tsallis has shown that a formalism of statistical mechanics can be consistently developed in terms of this generalized entropy \cite{Ts09}.

A consequence of the non- extensive formalism is that the distribution function which maximizes $S_q$ is non- Maxwellian \cite{Sil98,Lim00,Mu06}. For $q=1$, the Maxwell distribution function is reproduced. But for $q<1$, high energy states are more probable than in the extensive case. On the other hand, for $q>1$ high energy states are less probable than in the extensive case, and there is a cutoff beyond which no states exist.

\subsection{Maxwellian Distribution}

In stars, the thermonuclear reaction rate with a Maxwellian distribution is
given by \cite{Fow67}
\begin{eqnarray}
R_{ij}&=&\frac{N_{i}N_{j}}{1+\delta_{ij}} \langle \sigma v \rangle = \frac{N_{i}N_{j}}{1+\delta_{ij}}\left(\frac{8}{\pi\mu}\right)^{\frac{1}{2}}\left(\frac{1}{k_BT}\right)^{\frac{3}{2}}\nonumber \\
&\times&\int_{0}^{\infty}dES(E) \exp\left[-\left(\frac{E}{k_BT}+2\pi\eta(E)\right)\right], \label{rij}
\end{eqnarray}
where $\sigma$ is the fusion cross section, $v$ is the relative velocity of the $ij$-pair, $N_i$ is the number of nuclei of species $i$, $\mu$ is the reduced mass of $i+j$, $T$ is the temperature, $S(E)$ is the astrophysical S-factor, and $\eta=Z_iZ_je^2/\hbar v$ is the Sommerfeld parameter, with $Z_i$ the i-th nuclide charge and $E=\mu v^2/2$ is the relative energy of $i+j$. 

The energy dependence of the reaction cross sections is usually expressed in terms of the equation
\begin{equation}
 \sigma(E) = {S(E)\over E}\exp\left[-{2\pi\eta(E)}\right].
\label{se}
\end{equation} 
We write $2\pi \eta=b/\sqrt{E}$, where 
\begin{equation}
b=0.9898 Z_iZ_j\sqrt{A} \ {\rm MeV}^{1/2}, \label{b}
\end{equation} 
where $A$ is the reduced mass in amu. The factor $1+\delta_{ij}$ in the denominator of Eq. \eqref{rij} corrects for the double-counting when $i=j$. The S-factor has a relatively weak dependence on the energy $E$, except when it is close to a resonance, where it is strongly peaked.

\subsection{Non-Maxwellian Distribution}

The non-extensive description of the Maxwell-Boltzmann distribution corresponds to the substitution $f(E) \rightarrow f_q(E)$, where \cite{Ts09}
\begin{eqnarray}
f_q(E)&=&\left[1-\frac{q-1}{k_BT}E\right]^{\frac{1}{q-1}}\nonumber \\&\stackrel{\small q\rightarrow1}{\longrightarrow}&\exp\left(-\frac{E}{k_BT}\right),
\quad 0<E<\infty. \label{fene}
\end{eqnarray}
If $q-1<0$, Eq. \eqref{fene} is real for any value of $E\ge 0$. However,
if $q-1>0$, $f(E)$ is real only if the quantity in square brackets is positive. This means that
\begin{eqnarray}
0 \le E \le {k_BT\over q-1},&& \ \ \ \  {\rm if} \ \ q\ge 1 \nonumber \\
0\le E,&&  \ \ \ \  {\rm if} \ \ q\le 1.
\end{eqnarray}
Thus, in the interval $0<q<1$ one has $0<E<\infty$ and for $1<q<\infty$ one has
$0<E<E_{\mathrm{max}}={k_BT}/(q-1)$.

\begin{center}
\begin{figure}[t]
\includegraphics[width=85mm]{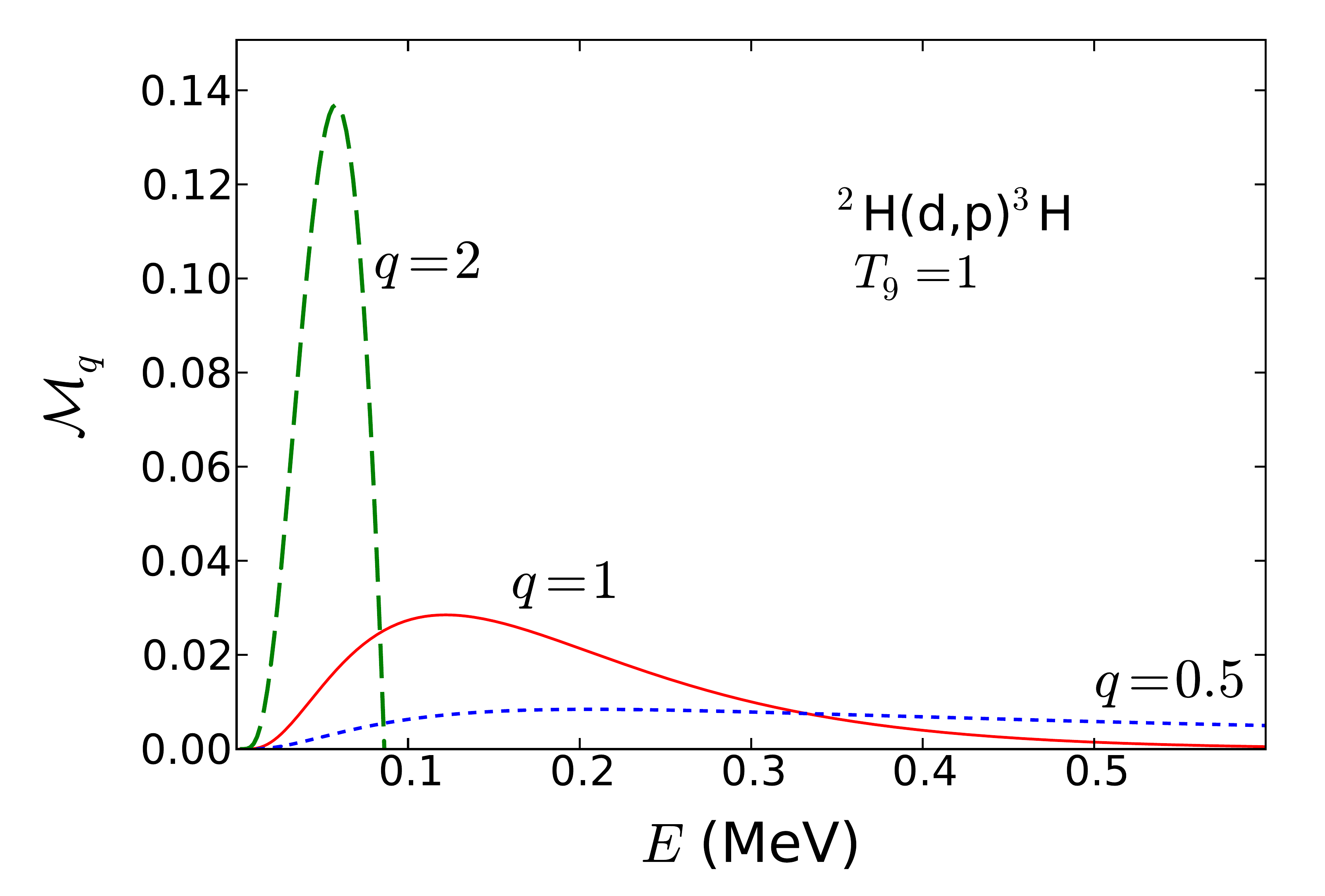}
\caption{Modified Gamow distributions  ${\cal M}_{q}(E,T)$ of deuterons relevant for the reaction $^2$H(d,p)$^3$H at $T_9=1$. The solid line, for $q=1$, corresponds to the use of a Maxwell-Boltzmann distribution. Also shown  are the results when using non-extensive distributions for $q=0.5$ (dotted line) and $q=2$ (dashed line). }\label{fmaxwell} 
\end{figure}
\end{center}

With this new statistics, the reaction rate becomes 
\begin{equation}
R_{ij}= \frac{N_{i}N_{j}}{1+\delta_{ij}} I_q, \label{rijne}
\end{equation}
and  the rate integral, $I_q$, is given by 
\begin{equation}
I_{q}=\int_{0}^{E_{max}}dES(E){\cal M}_{q}(E,T),\label{Iq}
\end{equation}
where the ``modified" Gamow energy distribution is 
\begin{eqnarray}
{\cal M}_{q}(E,T)&=&{\cal A}(q,T)\left(1-\frac{q-1}{k_BT}E\right)^{\frac{1}{q-1}}e^{-b/\sqrt{{E}}} \nonumber \\
&=&{\cal A}(q,T) \left(1-\frac{q-1}{0.08617T_9}E\right)^{\frac{1}{q-1}}\nonumber \\
&\times&\exp\left[-0.9898 Z_iZ_j\sqrt{A\over E}\right]
\label{MqET}\end{eqnarray}
is the non-extensive Maxwell velocity distribution, $E_{max}=\infty$ for $0<q<1$  and $E_{\mathrm{max}}={k_BT}/(1-q)$ for $1<q<\infty$, and $E$ in MeV units. ${\cal A} (q,T)$ is a normalization constant which depends on the temperature and the non-extensive parameter $q$.

\subsection{Non-Maxwellian distribution for relative velocities}
It is worthwhile to notice that if the  one-particle energy distribution is Maxwellian, it does not necessarily imply that the relative velocity distribution is also Maxwellian.   Additional assumptions about correlations between particles are necessary to deduce that the relative-velocity distribution, which is the relevant quantity for rate calculations, is also Maxwellian. This has been discussed in details in Refs. \cite{MQ05,Ts88,Kan05} where  non-Maxwellian distributions, such as in Eq. \eqref{MqET}, were shown to arise from non-extensive statistics.

Here we show that, if the  non-Maxwellian particle velocity distribution is given by Eq. \eqref{fene}, then a two-particle relative can be modified to account for the center of mass recoil.
Calling the kinetic energy of a particle $E_i$, this distribution is given by
\begin{equation}
f_q^{(i)} = \left(1 - \frac{q - 1}{kT}E_i\right)^{\frac{1}{q - 1}} \rightarrow \exp{\left[-\left(\frac{E_i}{kT}\right)\right]}\label{onepart}
\end{equation}

The two-particle energy distribution  is $f_q^{(1)}f_q^{(2)}$. We now exponentiate the Tsallis distribution.
\begin{equation}
f_q^{(i)} = \exp\left\{\frac{1}{q - 1} \left[\ln\left(1 - \frac{q - 1}{kT} E_i\right)\right]\right\}
\end{equation}
and the product $f_q^{(12)}=f_q^{(1)}f_q^{(2)}$ reduces to

\begin{equation}
f_q^{(12)} = \exp\left\{\frac{1}{q - 1} \left[\ln\left(1 - \frac{q - 1}{kT} E_1\right)\left(1 - \frac{q - 1}{kT}E_2)\right)\right]\right\}
\end{equation}

Since $E_i = {m_i v_i^2}/{2}$, and thus, $E_1 + E_2 = {\mu v^2}/{2} + {M V^2}/{2}$, where
$\mu$ is the reduced mass of the two particles, $M = m_1 + m_2$, $v$ is the relative velocity, and $V$ the center of mass velocity,  the product inside the natural logarithm can be reduced to
\begin{eqnarray}
&& 1 - \frac{1 - q}{kT} \left(\frac{\mu v^2}{2} + \frac{M V^2}{2}\right) + \left(\frac{1 - q}{kT}\right)^2 \frac{\mu v^2}{2}\frac{M V^2}{2} \nonumber \\
&&=\left(1 - \frac{1 - q}{kT} \frac{\mu v^2}{2}\right)\left(1 - \frac{1 - q}{kT} \frac{M V^2}{2}\right)\label{factoriz}
\end{eqnarray}

Thus, the two-body distribution factorizes into a product of relative and center of mass parts
\begin{eqnarray}
f_q^{(12)}(v,V,T) =  f_q^{(rel)}(v,T) f_q^{(cm)}(V,T)
\end{eqnarray}
where  
\begin{eqnarray}
f_q^{(rel)}(v,T)&=&{\cal A}_{rel}(q,T)\left(1 - \frac{1 - q}{kT} \frac{\mu v^2}{2}\right)^{\frac{1}{q - 1}}\nonumber\\
f_q^{(cm)}(V,T)&=&{\cal A}_{cm}(q,T)\left(1 - \frac{1 - q}{kT} \frac{M V^2}{2}\right)^{\frac{1}{q - 1}},
\end{eqnarray}
with the normalization constants obtained from the condition,
\begin{equation}
\int d^3v d^3Vf_q^{(12)}(v,V,T) = 1.
\end{equation}
Because the distribution factorizes, the unit normalization can be achieved for the relative and c.m. distributions separately.
The distribution needed in the reaction rate formula is, therefore,
\begin{equation}
f_q(v,T) = \int d^3V f_q^{(12)}(v,V,T)= f_q^{(rel)}(v,T),
\end{equation}
which attains the same form for as the absolute distribution.

In the limit $q \rightarrow 1$ the two-particle  distribution reduces to a Gaussian, with the last term in the left-hand-side of Eq. \eqref{factoriz} dropping out,
\begin{equation}
f_q^{(12)}= {\cal A}(q,T) \exp\left\{\left[-\left(\frac{{\mu v^2}/{2} + {MV^2}/{2}}{kT}\right)\right]\right\},
\end{equation}
as expected. 

\subsection{Equilibrium with electrons, photons and neutrinos}

One of the important questions regarding a plasma with particles (i.e., nuclei) described by the non-extensive statistics is how to generalize Fermi-Dirac, Bose-Einstein, and Tsallis statistics, to become more unified statistics with the distribution for the particles. This has been studied in Ref. \cite{BD00}, where it was shown that a similar non-extensive statistic for the  distribution can be obtained for fermions and bosons is given by 
\begin{equation}
n_q^{\pm} (E)=  {1\over 
\left[1-(q-1){(E-\mu)\over kT}\right]^{1\over q-1}\pm 1}, \label{nnon}
\end{equation}
where $\mu$ is the chemical potential. This reproduces the Fermi distribution, $n^+$, for $q\rightarrow 1$ and the Bose-Einstein distribution for photons, $n^-$, for $\mu=0$ and $q\rightarrow 1$. Planck's law for the distribution of radiation is obtained by multiplying $n^-$ in Eq. \eqref{nnon} (with $\mu=0$) by $\hbar \omega^2/(4\pi^2 c^2)$, where $E=\hbar \omega$. The number density of electrons can be obtained from $n^+$ in Eq. \eqref{nnon} with the proper phase-factors depending if the electrons are non-relativistic or relativistic. Normalization factors ${\cal A}^\pm (q,T)$ also need to be introduced, as before.

\begin{center}
\begin{figure}[t]
\includegraphics[width=85mm]{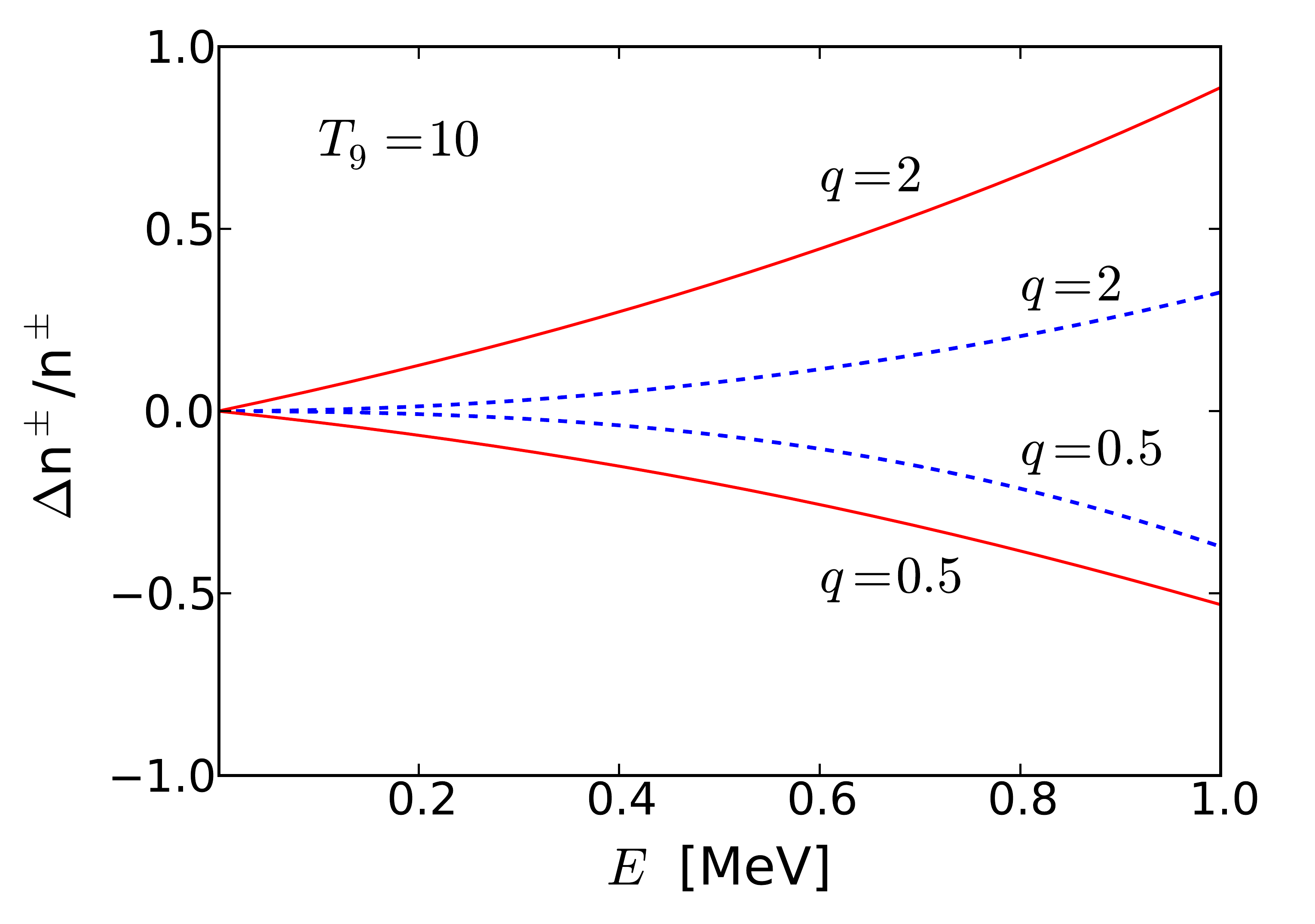}
\caption{The relative difference ratio $(n_q^{\pm}-n^\pm)/n^\pm$ between non-extensive, $n_q$, and extensive, $n=n_{q\rightarrow 1}$, statistics. Solid curves are for Fermi-Dirac statistics, $n^+$, and dashed curves are for Bose-Einstein statistics, $n^-$.  For both distributions, we use $\mu=0$.  Results are shown for $q=2$ and $q=0.5$, with $T_9=10$. }\label{fdn} 
\end{figure}
\end{center}

The electron density during the early universe varies strongly with the temperature. At $T_9=10$ the electron density is about $10^{32}$/cm$^3$, much larger than the electron number density at the center of the sun, $n_e^{sun}\sim 10^{26}$/cm$^3$.   The large electron density  is due to the $e^+e^-$ production by the abundant photons during the BBN.  However, the large electron densities do not influence the nuclear reactions during the BBN. In fact, the enhancement of the nuclear reaction rates due to electron screening were shown to be very small \cite{WBB11}. The electron Fermi energy for these densities are also much smaller than $kT$ for the energy relevant for BBN, so that one can also use $\mu=0$ in Eq. \eqref{nnon} for $n^+$.

In figure \ref{fdn} we plot the the relative difference ratio $(n_q^{\pm}-n^\pm)/n^\pm$ between non-extensive, $n_q$, and extensive, $n=n_{q\rightarrow 1}$, statistics. For both distributions, we use $\mu=0$. The solid curves are for Fermi-Dirac (FD) statistics, $n^+$, and dashed curves are for Bose-Einstein (BE) statistics, $n^-$. We show results for $q=2$ and $q=0.5$, with $T_9=10$. One sees that the non-extensive distributions are enhanced for $q>1$ and suppressed for $q<1$, as compared to the respective FD and BE quantum distributions. The deviations for the FD and BE statistics also grow larger with the energy. For example, the non-extensive electron distribution is roughly a factor 2 larger than the usual FD distribution at $E_e = 1$ MeV, at $T_9=10$. 

While we don't obtain numerical results with modified Fermi-Dirac and Bose-Einstein distributions here, it can be expected that these generalizations  will have a strong influence on the freezout temperature and the neutron to proton, n/p, ratio.   A numerical study of this problem may be presented in another paper.
The freezout temperature, occurs when the rate, $\Gamma \sim \left<\sigma v\right>$, for weak reaction $\nu_e + n \rightarrow p+e^-$ becomes slower than the expansion rate of the Universe. Because during the BBN  the densities of all particles, including neutrinos,  are low compared to $kT$, the chemical potential $\mu$ can be set to zero in the calculation of all reaction rates. The adoption of non-extensive quantum distributions such as in Eq. \eqref{nnon} will lead to the same powers of the temperature as those predicted by the FD distribution and the Bose-Einstein distribution. For example, Planck law for the total  blackbody is $U \propto T^4$, being form invariant with respect to non-extensivity entropic index $q$ which determines the the degree of non-extensivity \cite{BSD02}. This result means that the weak decay reaction rates do not depend on the non-extensive parameter $q$. The freezout temperature and n/p ratio remain the same as before.

More detailed studies have indicated that Planck's law of blackbody radiation and other thermodynamical quantities arising from non-extensive quantum statistics can yield  different powers of temperature than for the non-extensive case \cite{Ara03,Nau03,Tsa04,CDT04}. If that is the case, then a study of the influence of non-extensive statistics on the weak-decay rates and electromagnetic processes during BBN is worth pursuing.

\subsection{Thermodynamical equilibrium}
 
The  physical appeal for non-extensivity is the role of long-range interactions, which also implies non-equilibrium.
Accepting non-extensive entropy means abandoning the most important concept of thermodynamics, namely the tendency of any system to reach equilibrium. This also means that the concept of non-extensivity means renouncing to the second law of thermodynamics altogether! 

The comments above, which seem to be shared by part of the community (see, e.g., \cite{Nau03,ZM03,ZM04,BDR06,Dau07,Tou13}), are worrisome when one has to consider a medium composed of particles obeying classical and quantum statistics. It is not clear for example if the non-extensive parameter $q$ has to be the same for all particle distributions, both classical and quantum. Even worse is the possibility that the temperatures are not the same for the different particle systems in the plasma.

In the present work, we will avoid a longer discussion on the validity of the Tsallis statistics for a plasma such as that existing during the BBN. We will only consider the effect of its use for calculating nuclear reaction rates in the plasma, assuming that it can be described by a classical distribution of velocities. This study will allow us to constrain the non-extensive parameter $q$ based on a comparison with observations.

\section{Reaction rates during big bang nucleosynthesis}

Based on the abundant literature on non-extensive statistics (see, e.g., \cite{MQ05,HK08,Deg98,Cor99,Ts88}), 
we do not expect that the non-extensive parameter $q$ differs appreciably from the unity value.  However, in order to study the influence of a non-Maxwellian distribution on BBN we will explore values of $q$ rather different than the unity, namely, $q=0.5$ and $q=2$. This will allow us to pursue a better understanding of the nature of the physics involved in the departure from the BG statistics.  In figure \ref{fmaxwell} we plot the Gamow energy distributions of deuterons relevant for the reaction $^2$H(d,p)$^3$H at $T_9=1$. The solid line, for $q=1$, corresponds to the use of the  Maxwell-Boltzmann distribution. Also shown  are results for non-extensive distributions for $q=0.5$ (dotted line) and $q=2$ (dashed line). One observes that  for $q<1$, higher kinetic energies are more accessible than in the extensive case ($q=1$).  For $q>1$ high energies are less probable than in the extensive case, and there is a cutoff beyond which no kinetic energy is reached. In the example shown in the figure for $q=2$, the cutoff occurs at $0.086$ MeV, or $86$ keV.

\begin{center}
\begin{figure}[t]
\includegraphics[width=95mm]{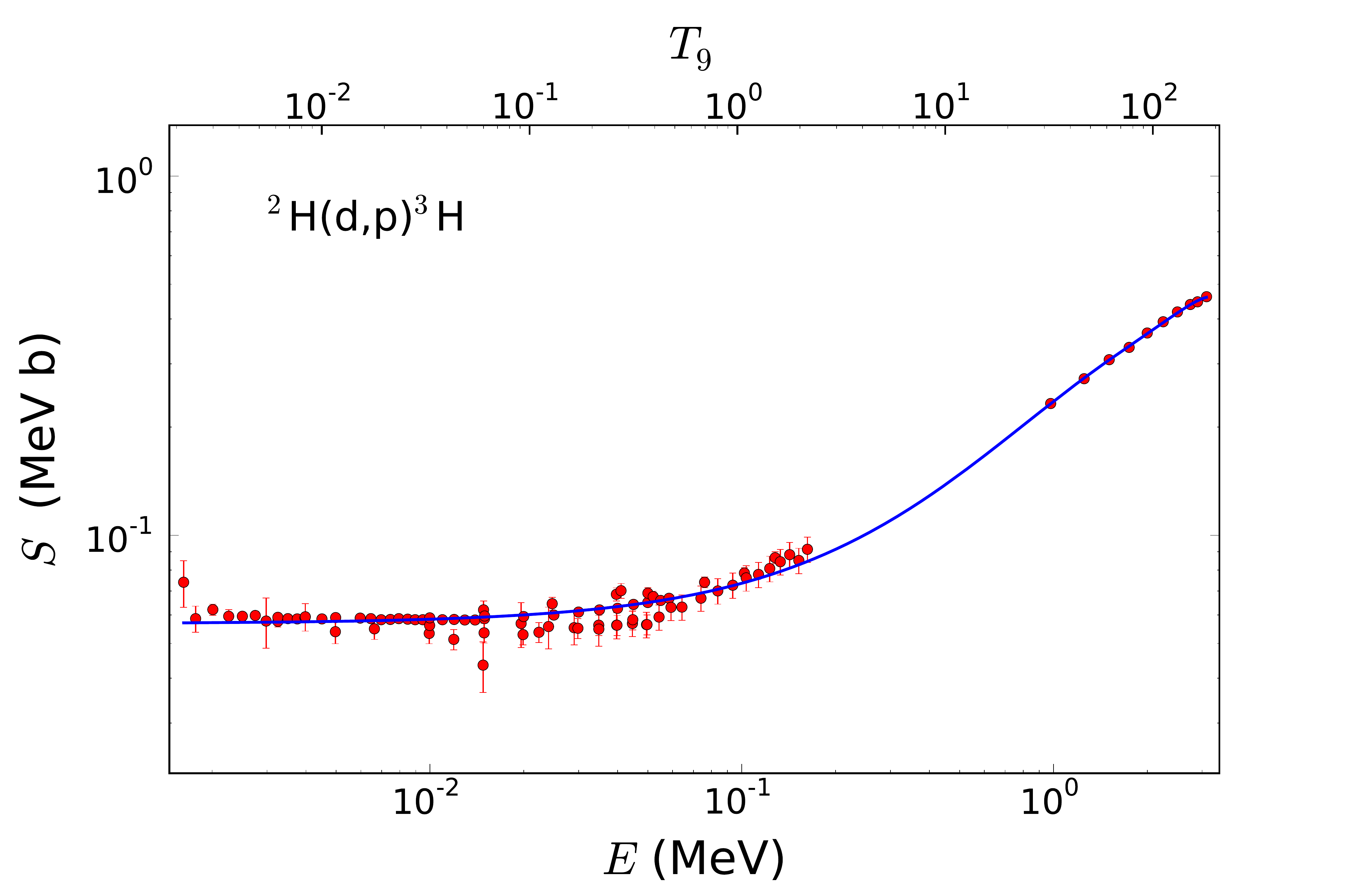}
\caption{S-factor for the reaction $^2$H(d,p)$^3$H as a function of the relative energy $E$ and of the temperature $T_9$. The data are from Refs. \cite{SC72,KR87,BR90,BO92,GR95}. The solid curve is a polynomial fit to the experimental data.}\label{sfddpt} 
\end{figure}
\end{center}

We will explore the modifications of the BBN elemental abundances  due to a variation of the non-extensive statistics parameter $q$. We will express our reaction rates in the form $N_A \langle \sigma v \rangle$ (in units of cm$^3$ mol$^{-1}$ s$^{-1}$), where $N_A$ is the Avogadro number and $\langle \sigma v \rangle$ involves the integral in Eq. \eqref{rij} with the Maxwell distribution $f(E)$ replaced by Eq. \eqref{fene}. First we show how the reaction rates are modified for $q\ne 1$.

In figure \ref{sfddpt} we show the S-factor for the reaction $^2$H(d,p)$^3$H as a function of the relative energy $E$.  Also shown is the dependence on $T_9$ (temperature in units of $10^9$ K)  for the effective Gamow energy 
\begin{equation}
E=E_0=0.122(Z_i^2Z^2_jA)^{1/3}T_9^{2/3} \ {\rm MeV}, \label{gamow}
\end{equation}
where $A$ is the reduced mass in amu.  The data are from Refs. \cite{SC72,KR87,BR90,BO92,GR95} and the solid curve is a chi-square polynomial function fit to the data.

\begin{center}
\begin{figure}[t]
\includegraphics[width=95mm]{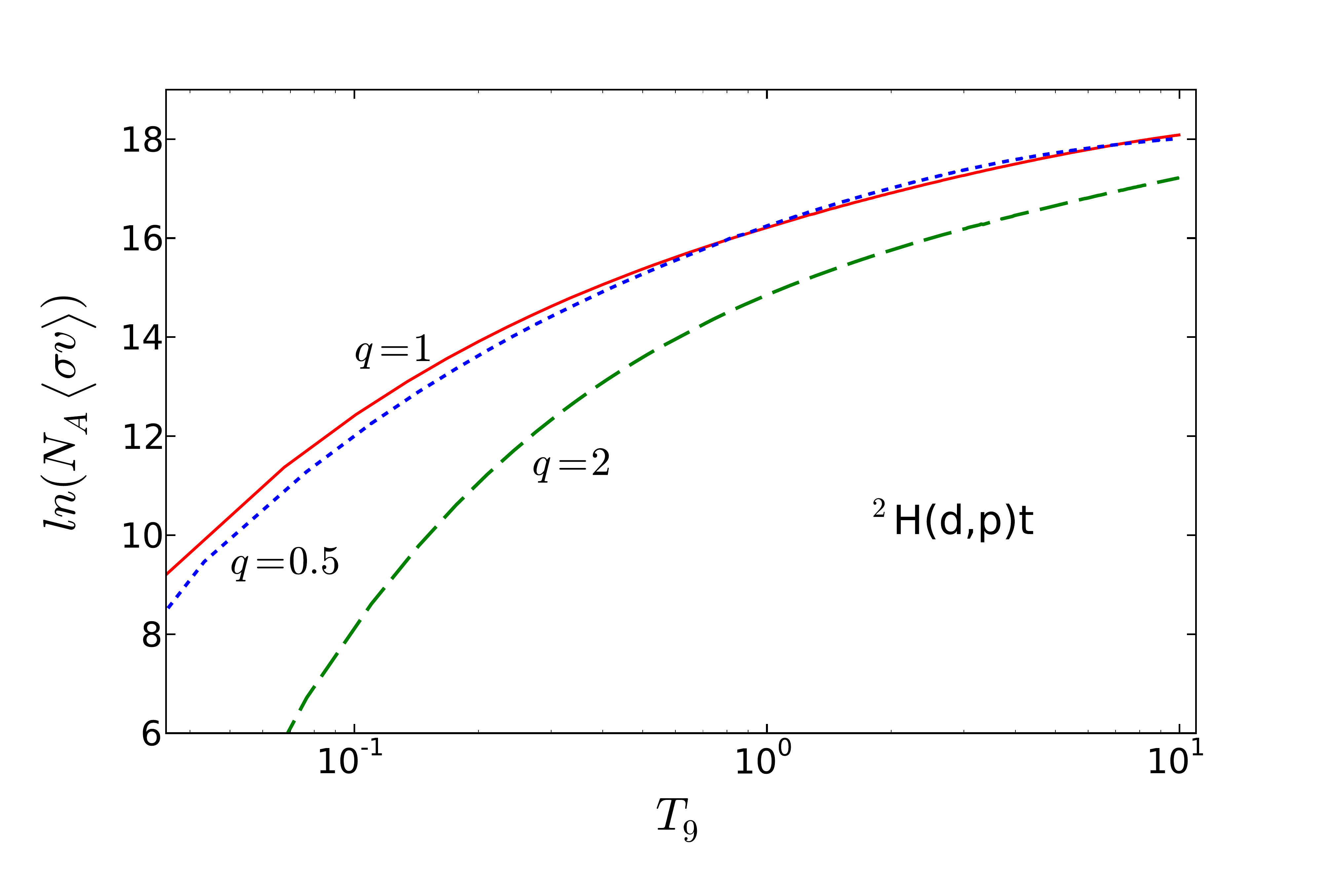}
\caption{Reaction rates for  $^2$H(d,p)$^3$H as a function of the temperature $T_9$ for different values of the non-extensive parameter $q$. The rates are given in terms of the natural logarithm of $N_A \langle \sigma v \rangle$ (in units of cm$^3$ mol$^{-1}$ s$^{-1}$).  Results with the use of non-extensive distributions for $q=0.5$ (dotted line) and $q=2$ (dashed line) are shown.}\label{rddpt} 
\end{figure}
\end{center}

Using the chi-square polynomial fit obtained to fit the data presented in figure \ref{sfddpt}, we show in figure \ref{rddpt} the reaction rates for  $^2$H(d,p)$^3$H as a function of the temperature $T_9$ for two different values of the non-extensive parameter $q$. The integrals in equation \eqref{Iq} are performed numerically. For charge particles, a good accuracy (witihin 0.1\%) is reached using the integration limits between  $E_0 - 5\Delta E$ and $E_0+5\Delta E$, where $\Delta E$ is given by Eq. \eqref{gamoww} below. The rates are expressed in terms of the natural logarithm of $N_A \langle \sigma v \rangle$ (in units of cm$^3$ mol$^{-1}$ s$^{-1}$). The solid curve corresponds to the usual Maxwell-Boltzmann distribution, i.e., $q=1$. The dashed and dotted curves are obtained for $q=2$ and $q=0.5$, respectively. In both cases, we see deviations from the Maxwellian rate. For $q>1$ the deviations are rather large and the tendency is an overall suppression of the reaction rates, specially at low temperatures.  This effect arises from the non-Maxwellian energy cutoff which for this reaction occurs at $0.086 T_9$ MeV and which prevents a great number of reactions to occur at higher energies. 

\begin{center}
\begin{figure}[h]
\includegraphics[width=95mm]{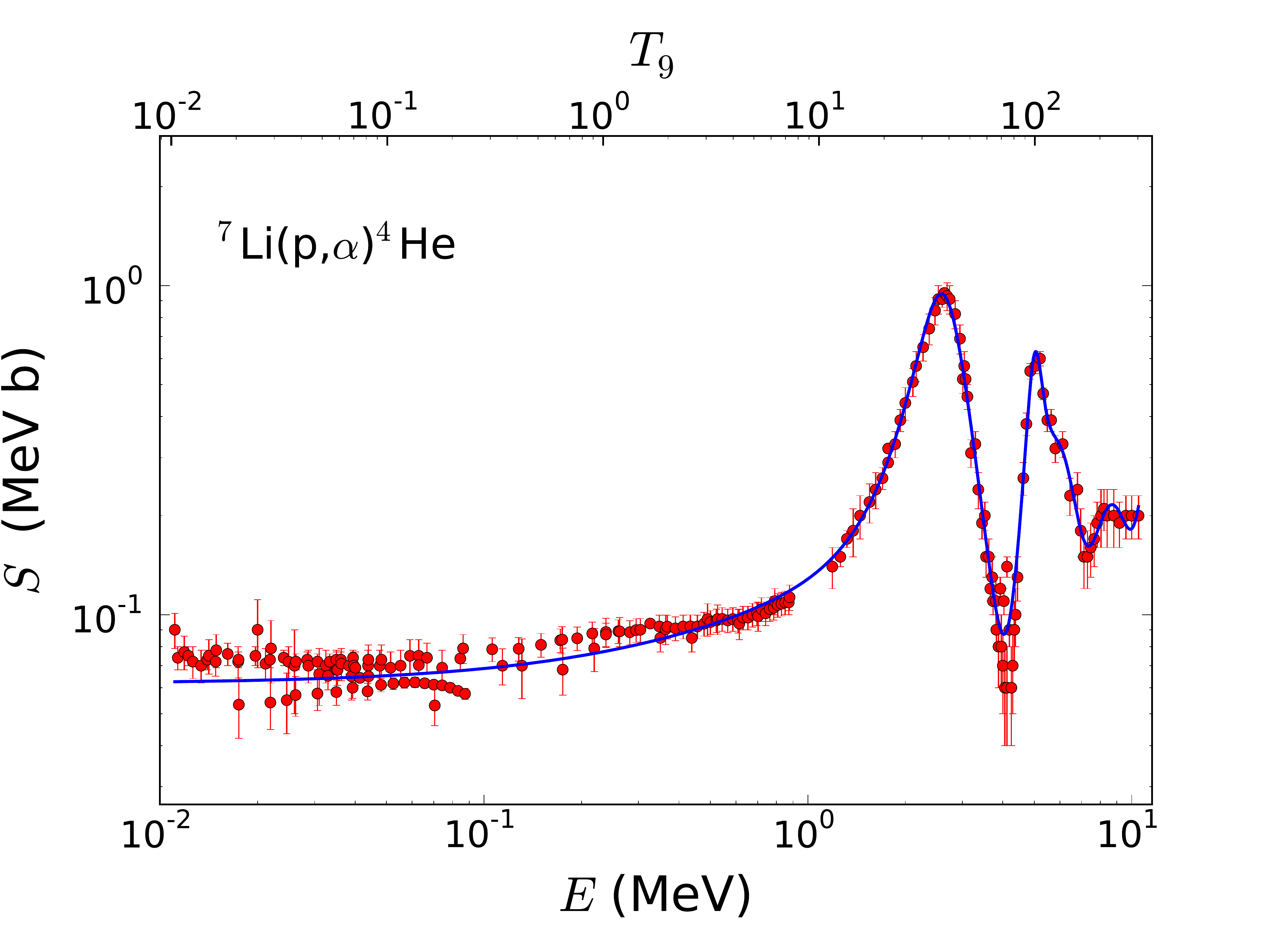}
\caption{S-factor for the reaction $^7$Li(p,$\alpha$)$^4$He as a function of the relative energy $E$ and of $T_9$. The data are from Refs. \cite{FR58,CA62a,CA62b,MA64,FI67,LE69,SP71,RO86,HA89,EN92a,EN92b}. The solid curve is a chi-square function fit to the data using a sum of polynomials plus Breit-Wigner functions.}\label{sf7liaa} 
\end{figure}
\end{center}

For $q<1$ the nearly similar result as with the Maxwell-Boltzmann distribution  is due to a competition between  suppression in reaction rates at low energies and their enhancement at high energies. The relevant range of energies is set by the Gamow energy which for a Maxwellian distribution is given by Eq. \eqref{gamow} and by the energy window,
\begin{equation}
\Delta E=0.2368(Z_i^2Z^2_jA)^{1/6}T_9^{5/6} \ {\rm MeV}, \label{gamoww}
\end{equation}
which for the reaction $^2$H(d,p)$^3$H amounts to $0.2368T_9^{5/6}$ MeV. This explains why, at $T_9=1$, the range of relevant energies for the calculation of the reaction rate is shown by the solid curve in figure \ref{fmaxwell}. For $q<1$ the Gamow window $\Delta E$ is larger and there is as much a contribution from the suppression of reaction rates at low energies compared to the Maxwell-Boltzmann distribution, as there is a corresponding enhancement at higher energies. This explains the nearly equal results shown in figure \ref{rddpt}  for $q=1$ and $q<1$. This finding applies to all charged particle reaction rates, except for those when the S-factor has a strong dependence on energy at, and around, $E=E_0$.  But no such behavior exists for the most important charged induced reactions in the BBN (neutron-induced reactions will be discussed separately).

\begin{center}
\begin{figure}[tbh]
\includegraphics[width=95mm]{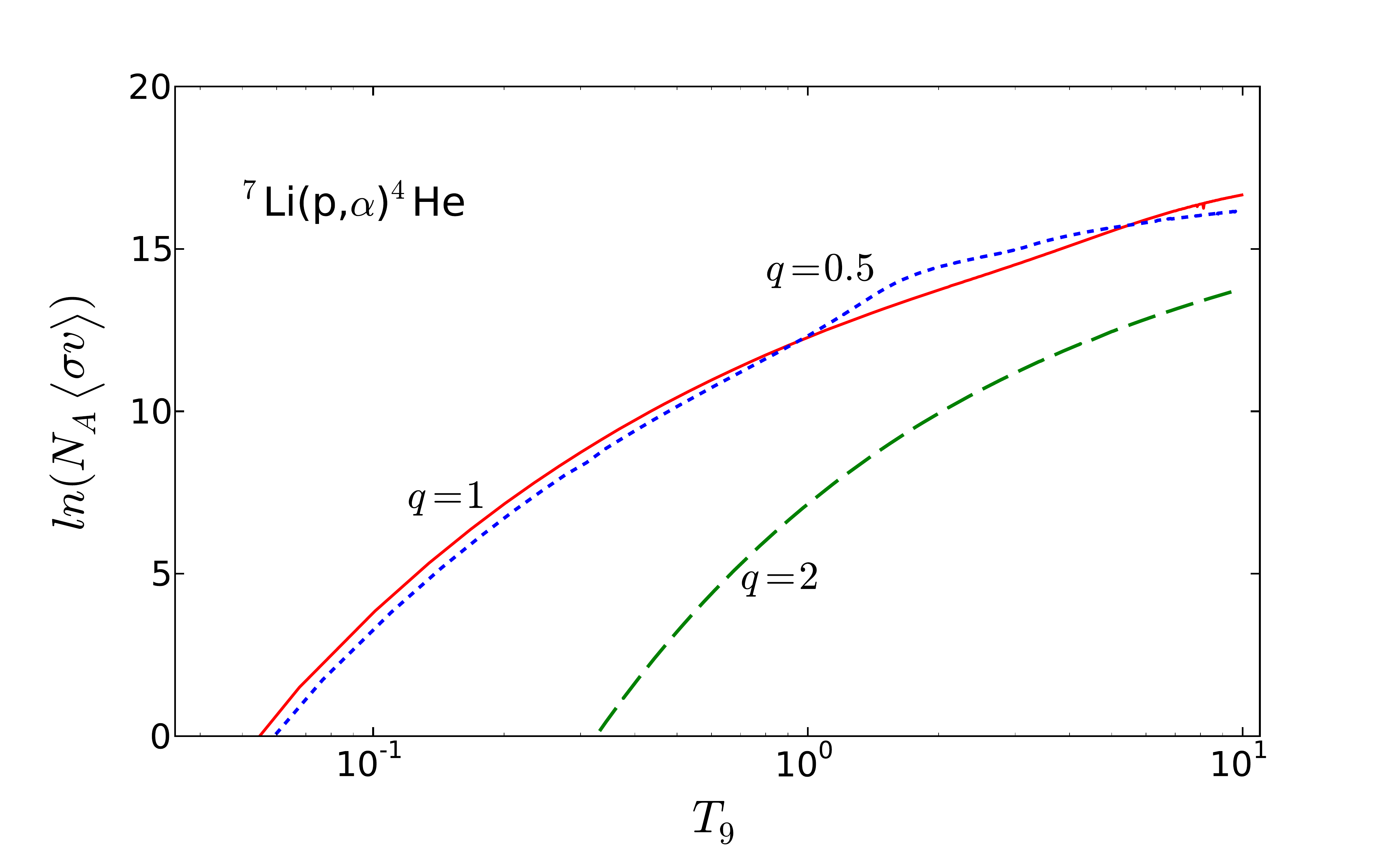}
\caption{Reaction rates for  $^7$Li(p,$\alpha$)$^4$He as a function of the temperature $T_9$ for two different values of the non-extensive parameter $q$. The rates are given in terms of the natural logarithm of $N_A \langle \sigma v \rangle$ (in units of cm$^3$ mol$^{-1}$ s$^{-1}$). Results with the use of non-extensive distributions for $q=0.5$ (dotted line) and $q=2$ (dashed line) are shown.}\label{r7liaa} 
\end{figure}
\end{center}

The findings described above for the reaction  $^2$H(d,p)$^3$H are not specific but apply to all charged particles of relevance to the BBN. We demonstrate this with one more example: the $^7$Li(p,$\alpha$)$^4$He reaction, responsible for $^7$Li destruction. 
In figure \ref{sf7liaa} we show the S-factor for this reaction  as a function of the relative energy $E$. One sees prominent resonances at higher energies.  Also shown in the figure is the dependence of the reaction on $T_9$. The data are from Refs. \cite{FR58,CA62a,CA62b,MA64,FI67,LE69,SP71,RO86,HA89,EN92a,EN92b} and the solid curve is a chi-square function fit to the data using a sum of polynomials plus Breit-Wigner functions.

Using the chi-square function fit obtained to fit the data presented in figure \ref{sf7liaa}, we show in figure \ref{r7liaa} the reaction rates for  $^7$Li(p,$\alpha$)$^4$He as a function of the temperature $T_9$ for two different values of the non-extensive parameter $q$. The rates are given in terms of the natural logarithm of $N_A \langle \sigma v \rangle$ (in units of cm$^3$ mol$^{-1}$ s$^{-1}$). The solid curve corresponds to the usual Maxwell-Boltzmann distribution, i.e., $q=1$. The dashed and dotted curves are obtained for $q=2$ and $q=0.5$, respectively. As with the reaction presented in figure \ref{rddpt},  in both cases we see deviations from the Maxwellian rate. But, as before, for $q=2$ the deviations are larger and the tendency is a strong suppression of the reaction rates as the temperature decreases.  It is interesting to note that the non-Maxwellian rates for $q=0.5$ are more sensitive to the resonances than for $q>1$. This is because, as seen in figure \ref{fmaxwell}, for $q<1$ the velocity distribution is spread to  considerably larger values of energies, being therefore more sensitive to the location of high energy resonances.  
 
\begin{center}
\begin{figure}[tbh]
\includegraphics[width=90mm]{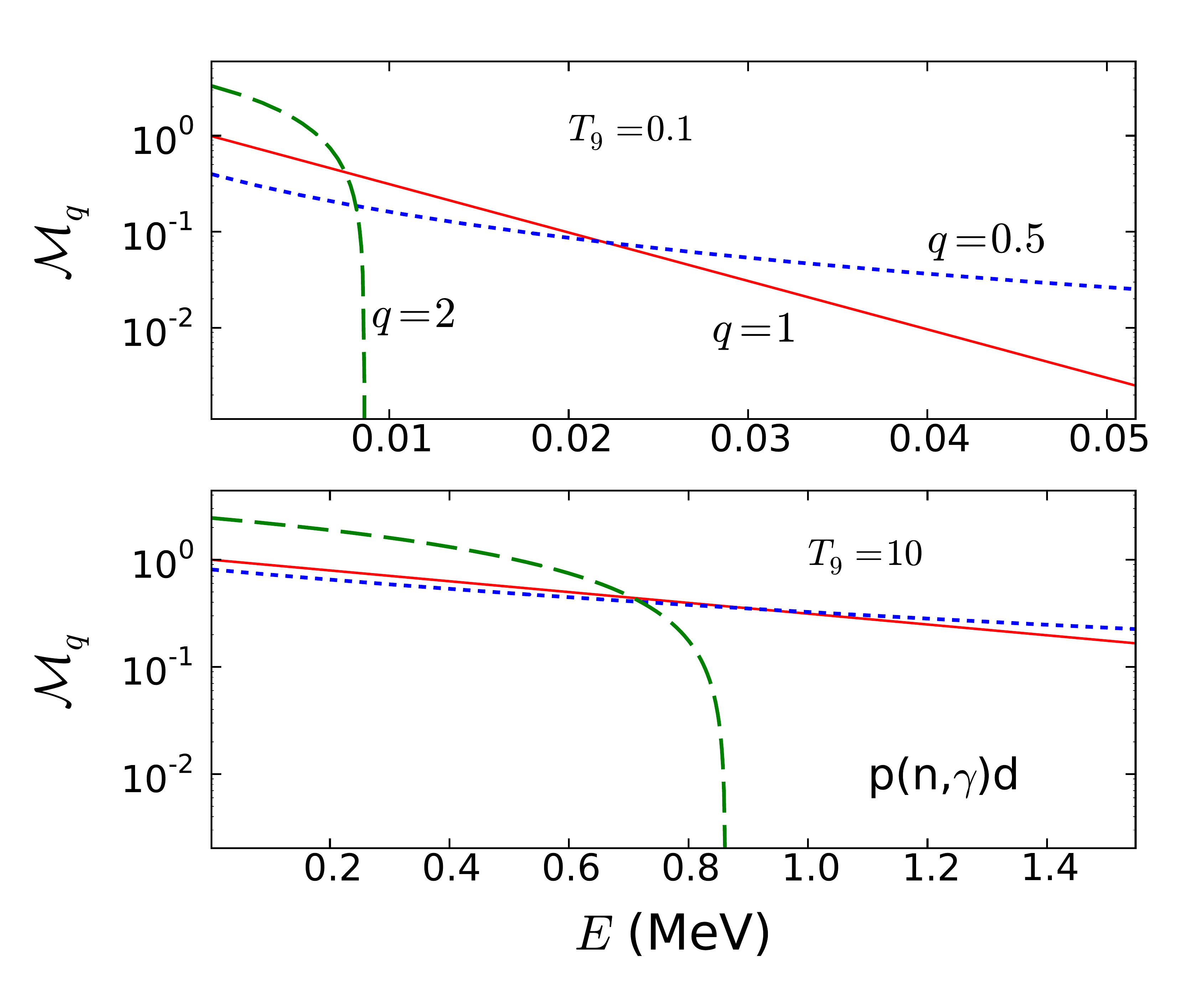}
\caption{Spectral function  ${\cal M}_{q}(E,T)$ for protons and neutrons relevant for the reaction p(n,$\gamma$)d at $T_9=0.1$ (upper panel) and $T_9=10$ (lower panel). The solid line, for $q=1$, corresponds to the usual Boltzmann distribution. Also shown  are non-extensive distributions for $q=0.5$ (dotted line) and $q=2$ (dashed line). }\label{png} 
\end{figure}
\end{center}

We now turn to neutron induced reactions, which are only a few cases of high relevance for the BBN, notably the p(n,$\gamma$)d, ${^3}$He(n,p)t, and $^7$Be(n,p)$^7$Li reactions. For neutron induced reactions, the cross section at low energies is usually proportional to $1/v$, where $v=\sqrt{2mE}/\hbar$ is the neutron velocity. Thus, it is sometimes appropriate to rewrite Eq. \eqref{se} as
\begin{equation}
 \sigma(E) = {S(E)\over E}= {R(E) \over \sqrt{E}}
\label{sen}
\end{equation}
where  $R(E)$ is a slowly varying function of energy  similar to an $S$-factor. The distribution function within the reaction rate integral \eqref{Iq} is also rewritten as
\begin{eqnarray}
{\cal M}_{q}(E,T)&={\cal A} (q,T)f_q(E)=&{\cal A} (q,T)\left(1-\frac{q-1}{k_BT}E\right)^{\frac{1}{q-1}}. \nonumber \\
\label{MqET2}\end{eqnarray}
The absence of the tunneling factor $\exp(-b/\sqrt{E})$ in Eq. \eqref{MqET2} inhibts the dependence of the reaction rates on the non-extensive parameter $q$.

In figure \ref{png} we plot the kinetic energy distributions of nucleons relevant for the reaction p(n,$\gamma$)d at $T_9=0.1$ (upper panel) and $T_9=10$ (lower panel). The solid line, for $q=1$, corresponds to the usual Boltzmann distribution. Also shown  are results for the non-extensive distributions for $q=0.5$ (dotted line) and $q=2$ (dashed line). One observes that, as for the charged particles case,  with $q<1$ higher kinetic energies are more probable than in the extensive case ($q=1$).  With $q>1$ high energies are less accessible than in the extensive case, and there is a cutoff beyond which no kinetic energy is reached.  A noticeable difference form the case of charged particles is the absence of the Coulomb barrier and a correspondingly lack of suppression of the reaction rates at  low energies. As the temperature increases, the relative difference between the  Maxwell-Boltzmann and the non-Maxwellian distributions decrease  appreciably. This will lead to a rather distinctive pattern of the reaction rates for charged compared to neutron induced reactions.

\begin{center}
\begin{figure}[t]
\includegraphics[width=85mm]{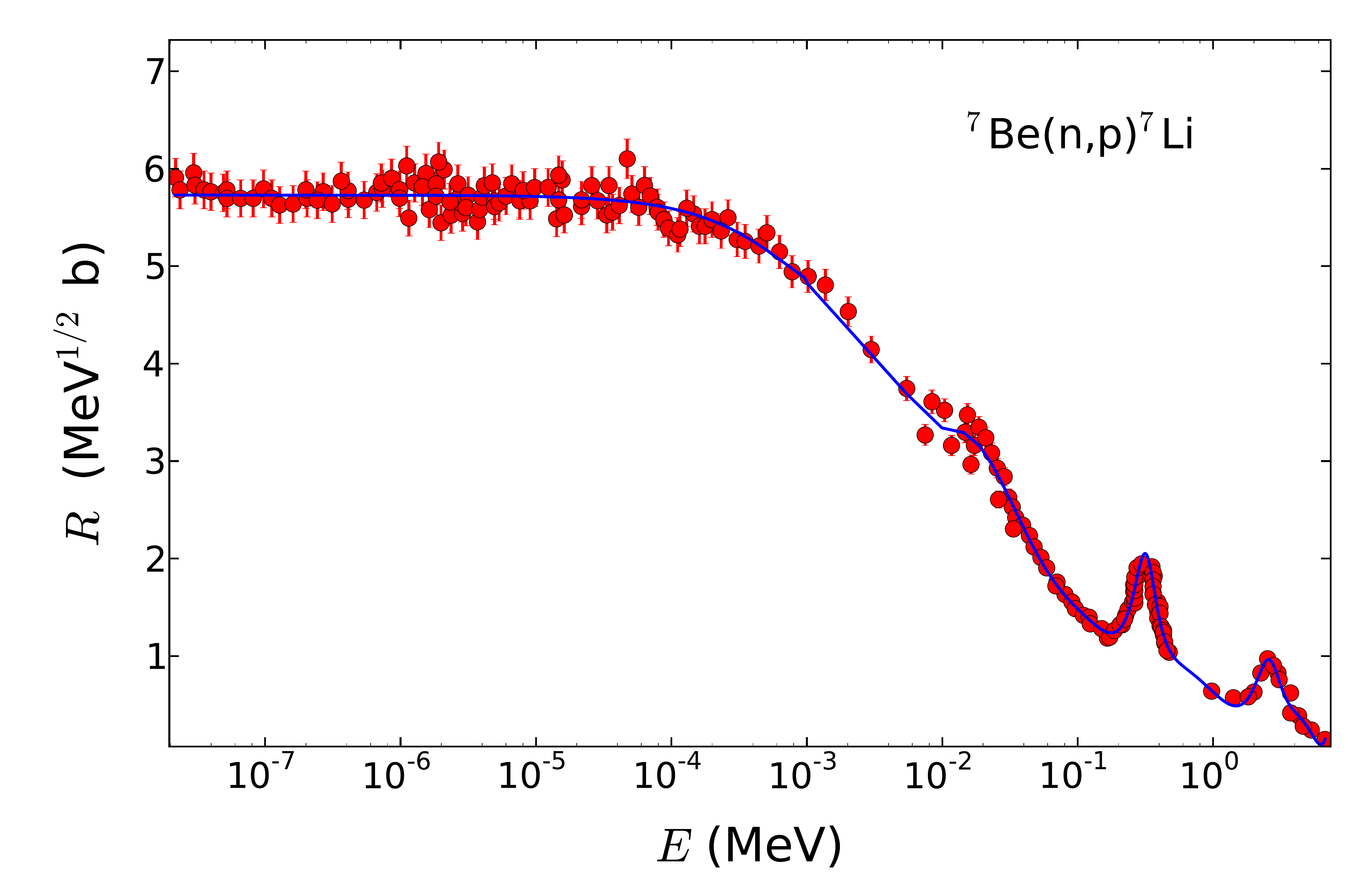}
\caption{The energy dependence of $R(E)=S(E)\sqrt{E}$ for the reaction $^7$Be(n,p)$^7$Li. is shown in figure \ref{nBep}. The experimental data were collected from Refs. \cite{GM59,BP63,Sek76,Pop76,Koe88}. The solid curve is a function fit to the experimental data  using a set of polynomials and Breit-Wigner functions.}\label{nBep} 
\end{figure}
\end{center}

For neutron-induced reactions, a good accuracy (within 0.1\%) for the numerical calculation of the reaction rates with Eq. \eqref{Iq} is reached using the integration limits between  $E = 0$ and $E = 20k_BT$. As an example we will now consider  the reaction $^7$Be(n,p)$^7$Li. The energy dependence of $R(E)=S(E)\sqrt{E}$ for this reaction is shown in figure \ref{nBep}. The experimental data were collected from Refs. \cite{GM59,BP63,Sek76,Pop76,Koe88}.

\begin{center}
\begin{figure}[t]
\includegraphics[width=85mm]{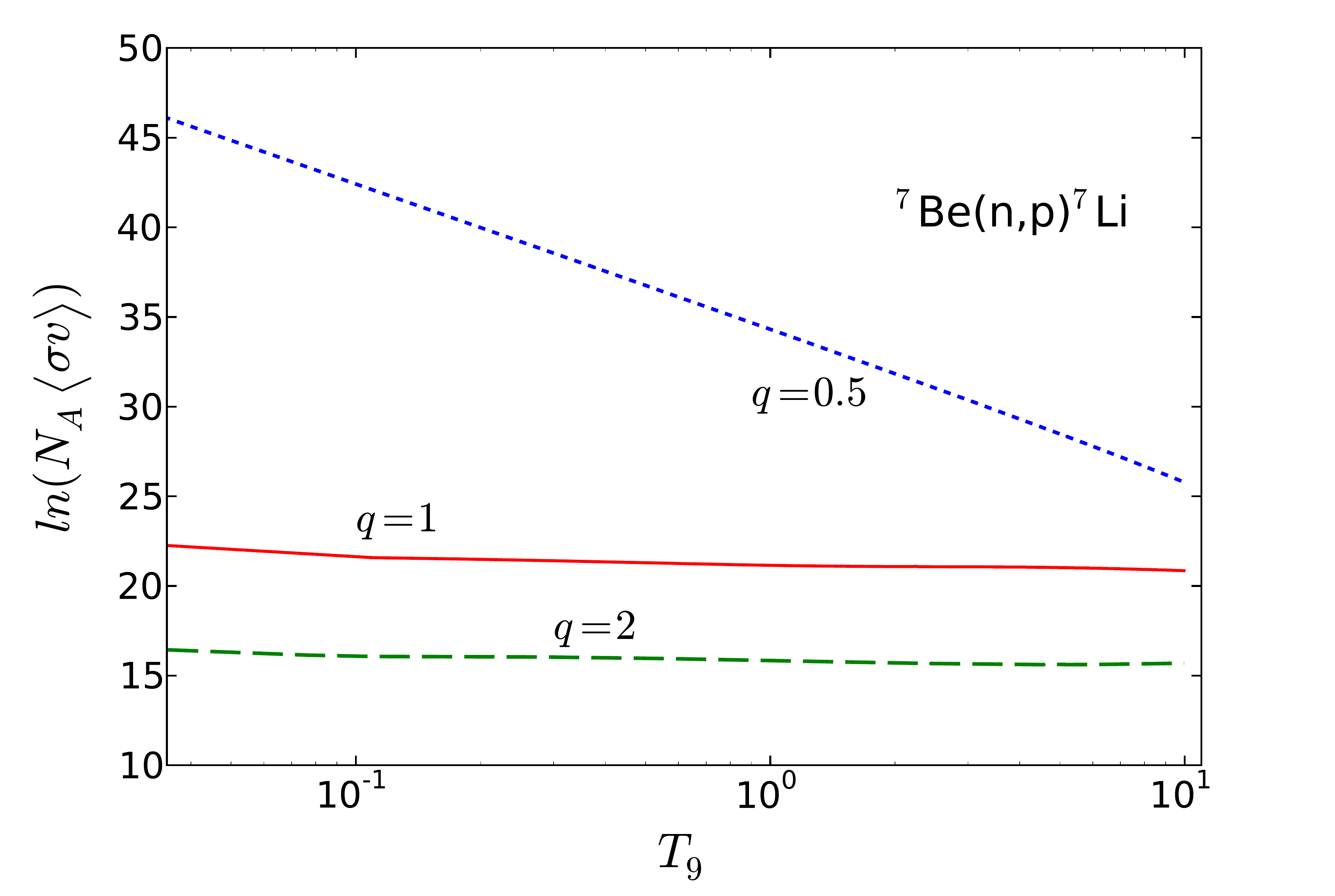}
\caption{Reaction rates for  $^7$Be(n,p)$^7$Li as a function of the temperature $T_9$ for two different values of the non-extensive parameter $q$. The rates are given in terms of the logarithm of $N_A \langle \sigma v \rangle$ (in units of cm$^3$ mol$^{-1}$ s$^{-1}$). Results with the use of non-extensive distributions for $q=0.5$ (dotted line) and $q=2$ (dashed line) are shown.}\label{nBep2} 
\end{figure}
\end{center}

Using the chi-square fit with a sum of polynomials and Breit-Wigners obtained to reproduce the data in figure \ref{nBep}, we show in figure \ref{nBep2} the reaction rates for  $^7$Be(n,p)$^7$Li as a function of the temperature $T_9$ for different values of the non-extensive parameter $q$. The rates are given in terms of the natural logarithm of $N_A \langle \sigma v \rangle$ (in units of cm$^3$ mol$^{-1}$ s$^{-1}$). The solid curve corresponds to the usual Boltzmann distribution, i.e., $q=1$. The dashed and dotted curves are obtained for $q=2$ and $q=0.5$, respectively. In contrast to reactions induced by charged particles, we now see strong deviations from the Maxwellian rate both for $q>1$ and $q<1$. For $q<1$ the deviations are larger at small temperatures and decrease as the energy increase, tending asymptotically to the Maxwellian rate at large temperatures. This behavior can be understood from figure \ref{png} (for $^7$Be(n,p)$^7$Li the results are nearly the same as in Fig. \ref{png}). At small temperatures, e.g. $T_9=0.1$,  the  distribution for $q=0.5$ is strongly enhanced at large energies and the tendency is that the reaction rates increase at low temperatures. This enhancement disappears as the temperature increase (lower panel of figure \ref{png}). For $q=2$ the reaction rate is suppressed, although not as much as for the charged-induced reactions, the reason being due a compensation by an increase because of normalization at low energies. 

Having discussed the dependence of the reaction rates on the non-extensive parameter $q$ for a few standard reactions, we now consider the implications of the non-extensive statistics to the predictions of the BBN. It is clear from the results presented above that an appreciable impact on the abundances of light elements will arise.

\section{BBN with non-extensive statistics} 

The BBN is sensitive to certain parameters, including the baryon-to-photon ratio, number of neutrino families, and the
neutron decay lifetime. We use the values $\eta = 6.19 \times 10^{-10}$, $N_\nu = 3$, and $\tau_n = 878.5$ s for the baryon-photon ratio, number of neutrino families, and neutron-day lifetime, respectively. Our BBN abundances were calculated with a modified version of the standard BBN code derived from Refs.  \cite{WFH67,Kaw68,Kaw92}.

Although BBN nucleosynthesis  can involve reactions up to the CNO cycle \cite{Alan11},   the  most important  reactions which can significantly affect the predictions of the abundances of the light elements [${^4}$He, D, ${^3}$He, ${^7}$Li]  are n-decay, p(n,$\gamma$)d, d(p,$\gamma){^3}$He, d(d,n)${^3}$He, d(d,p)t, ${^3}$He(n,p)t, t(d,n)${^4}$He, ${^3}$He(d,p)${^4}$He, ${^3}$He$(\alpha,\gamma){^7}$Be, t$(\alpha,\gamma){^7}$Li, ${^7}$Be(n,p)${^7}$Li and ${^7}$Li(p,$\alpha){^4}$He. Except for these reactions, we have used the reaction rates needed for the remaining reactions from a compilation by NACRE \cite{Nacre} and that reported in Ref. \cite{Des04}. For the 11 reactions mentioned above, we have collected data from Refs. \cite{SKM93,Nacre,Des04}, and references mentioned therein (data for n(p,$\gamma$,d) reaction was taken from the on-line ENDF database \cite{ENDF} - see also \cite{Cyb04,And06}), fitted the S-factors with a sum of polynomials and Breit-Wigner functions and calculated the reaction rates for Maxwellian and non-Maxwellian distributions.

\subsection{Elemental abundances}

\begin{center}
\begin{figure}[t]
\includegraphics[width=85mm]{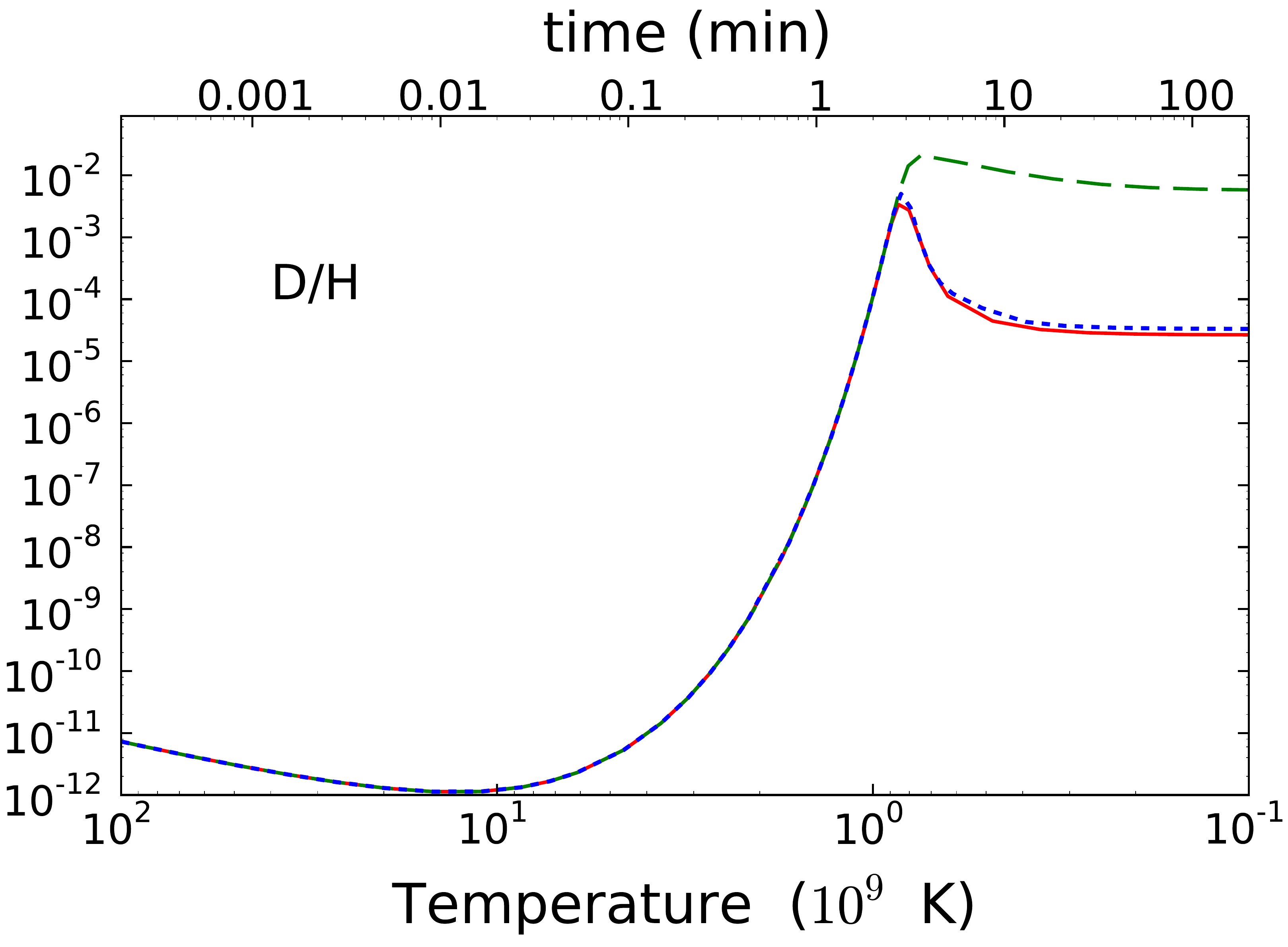}
\caption{Deuterium abundance. The solid curve is the result obtained with the standard Maxwell distributions for the reaction rates. Results with the use of non-extensive distributions for $q=0.5$ (dotted line) and $q=2$ (dashed line) are shown.}\label{abundd} 
\end{figure}
\end{center}

In figure \ref{abundd} we show the calculated deuterium abundance. The solid curve is the result with the standard Maxwell distributions for the reaction rates. Using the non-extensive distributions yields the dotted line for $q=0.5$  and the dashed line for $q=2$. It is interesting to observe that the deuterium abundances are only moderately modified due to the use of the non-extensive statistics for $q=0.5$.   Up to temperatures of the order of $T_9=1$, the abundance for D/H tends to agree for the extensive and non-extensive statistics. This is due to the fact that any deuterium that is formed is immediately destroyed (a situation known as the deuterium bottleneck). But, as the temperature decreases, the reaction rates for the p(n,$\gamma$)d reaction are considerably enhanced for $q=2$ (see figure \ref{png}), and perhaps more importantly, they are strongly suppressed for all other reactions involving deuterium destruction, as clearly seen in figure \ref{rddpt}. This creates an unexpected over abundance of deuterons for the non-extensive statistics with $q=2$. The deuterium, a very fragile isotope,   is easily destroyed after the BBN and astrated.  Its primordial abundance is determined from observations of interstellar clouds at high redshift, on the line of sight of distant quasars. These observations are scarce but allow to set an average value of ${\rm D/H} = 2.82^{ + 0.20}_{ - 0.19} \times 10^{-5}$ \cite{Pet08}. 
The predictions for the D/H ratio with the $q=2$ statistics (D/H = $5.70\times 10^{-3}$)   is about a factor 200 larger than those from the standard BBN model,  clearly in disagreement with the observation.
         
\begin{table}[htbp]
\vspace{0.0cm}
\centering
\caption{\label{tab:table1} Predictions of the BBN (with $ \eta_{WMAP}=6.2\times 10^{-10}$) with Maxwellian and non-Maxwellian distributions compared with observations. All numbers have the same power of ten as in the last column.}
\begin{tabular}{ccccccc}
\hline
\hline
 &Maxwell&Non-Max.&Non-Max.&Observation \\ 
 &   BBN  &   $q=0.5$             &$q=2$&\\ \hline

${^4}$He/H&0.249&0.243&0.141&$0.2561 \pm 0.0108$ \\
D/H&2.62&3.31&570&$2.82^{+0.20}_{-0.19}$($\times 10^{-5}$) \\
${^3}$He/H&0.98&0.91&69.1&$(1.1\pm 0.2)$($\times 10^{-5}$) \\
${^7}$Li/H&4.39&6.89&356.&$(1.58\pm 0.31) (\times 10^{-10})$ \\ \hline
\hline
\end{tabular} 
\vspace{0.0cm}
\label{tab1}
\end{table}

\begin{center}
\begin{figure}[t]
\includegraphics[width=90mm]{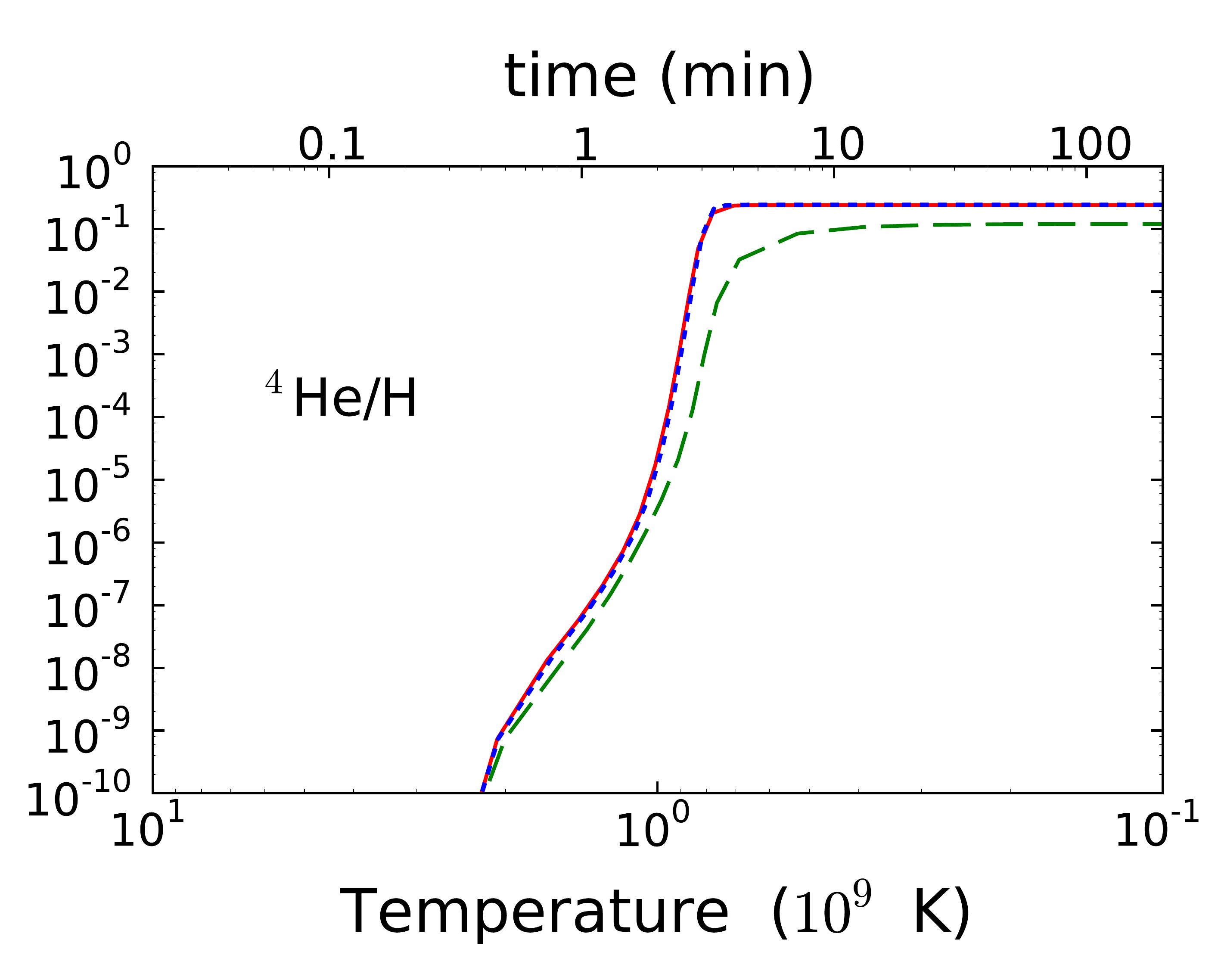}
\caption{$^4$He abundance. The solid curve is the result obtained with the standard Maxwell distributions for the reaction rates. Results with the use of non-extensive distributions for $q=0.5$ (dotted line) and $q=2$ (dashed line) are also shown.}\label{abundhe} 
\end{figure}
\end{center}

A much more stringent constraint for elemental abundances is given by $^4$He, which observations set at about $^4{\rm He/H} \equiv Y_p = 0.2561 \pm 0.0108$  \cite{Boe85,Ave10,IT10}. The $^4$He abundance generated from our BBN calculation is plotted in figure \ref{abundhe}. The solid curve is the result obtained with the standard Maxwell distribution for the reaction rates. Using the non-extensive distributions yields the dotted line for $q=0.5$  and the dashed line for $q=2$. Again, the predicted abundances for $q=2$ deviate substantially from standard BBN results. This time only about half  of $^4$He is produced with the use of a non-extensive statistics with $q=2$. The reason for this is the suppression of the reaction rates for formation of $^4$He with $q=2$ through the charged particle reactions  t(d,n)${^4}$He, ${^3}$He(d,p)${^4}$He.

\begin{center}
\begin{figure}[t]
\includegraphics[width=85mm]{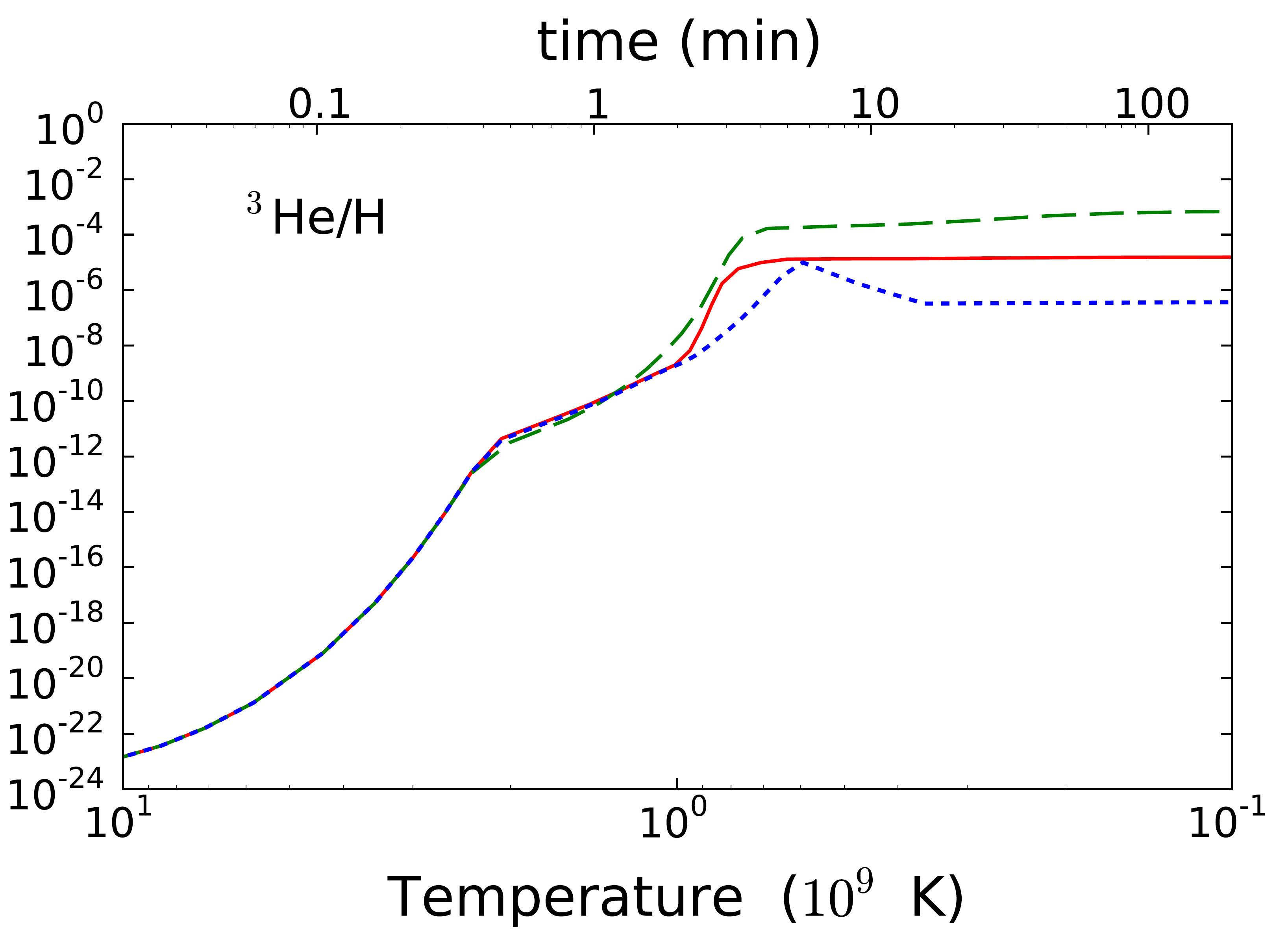}
\caption{$^3$He abundance. The solid curve is the result obtained with the standard Maxwell distributions for the reaction rates. Results with the use of non-extensive distributions for $q=0.5$ (dotted line) and $q=2$ (dashed line) are also shown.}\label{abundhe3} 
\end{figure}
\end{center}

A strong impact of using non-extensive statistics for both $q=0.5$ and $q=2$ values of the non-extensive parameter is seen in figure \ref{abundhe3} for the $^3$He abundance. While for $q=2$ there is an overshooting in the production of $^3$He, for $q=0.5$ one finds a smaller value than the one predicted by the standard BBN. This is due to the distinct results for the destruction of $^3$He through the reaction ${^3}$He(n,p)t, which is enhanced for $q=0.5$ and suppressed for $q=2$, in the same way as it happens for the reaction ${^7}$Be(n,p)${^7}$Li, shown in figure \ref{nBep}. $^3$He  is both produced and destroyed in stars and its abundance is still subject to large uncertainties, $^3{\rm He/H} = (1.1\pm 0.2) \times 10^{-5}$  \cite{Ban02,Van03}.

\begin{center}
\begin{figure}[t]
\includegraphics[width=90mm]{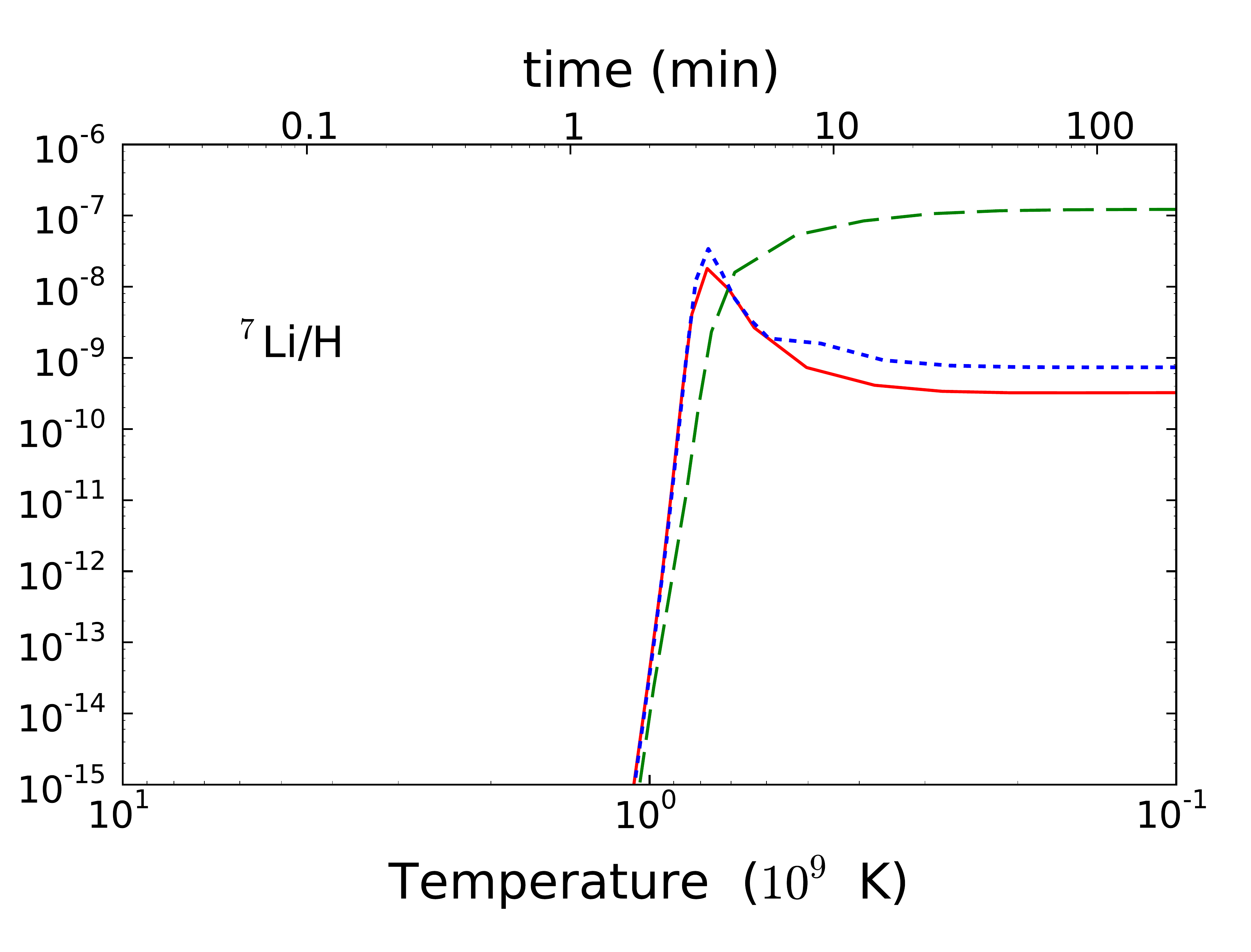}
\caption{$^7$Li abundance. The solid curve is the result obtained with the standard Maxwell distributions for the reaction rates. Results with the use of non-extensive distributions for $q=0.5$ (dotted line) and $q=2$ (dashed line) are also shown.}\label{abundli} 
\end{figure}
\end{center}

Non-extensive statistics for both $q=0.5$ and $q=2$ values also alter substantially the $^7$Li abundance, as shown in figure \ref{abundli}.  For both values of the non-extensive parameter  $q=2$, and $q=0.5$, there is an overshooting in the production of $^7$Li. The increase in the $^7$Li abundance is more accentuated for $q=2$. The lithium problem is associated with a smaller value of the observed $^7$Li abundance as compared the predictions of BBN. There has been many attempts to solve this problem by testing all kinds of modifications of the parameters of the BBN or the physics behind it (a sample of this literature is found in Refs. \cite{Fie11,Cyb04,NB00,BC08,BP09,Boy10,WBB11,KD11,FP12} and references therein). In the present case, the use of a non-Maxwellian velocity distribution seems to worsen this scenario. A recent analysis yields the observational value of 
$^7{\rm Li/H} = (1.58\pm 0.31) \times 10^{-10}$ \cite{Sb10}. 

\subsection{Sensitivity study}

We have calculated a window of opportunity for the non-extensive parameter $q$ with which one can reproduce the observed abundance of light elements. We chose the data for the abundances, $Y_i$, of $^4$He/H, D/H, $^3$He/H and $^7$Li as reference. We then applied the ordinary $\chi^2$ statistics, defined by the minimization of
\begin{equation}
\chi^2 = \sum_i {[Y_i(q)-Y_i(obs)]^2\over \sigma_i},
\end{equation}
where $Y_i(q)$ are the abundances obtained with the non-extensive statistics with parameter $q$, $Y_i(obs)$ are the observed abundances, and $\sigma_i$ the errors for each datum, and the sum is over all data mentioned in Table \ref{tab1}. From this chi-square fit  we conclude that  $q=1^{+0.05}_{-0.12}$ is compatible with observations.

No attempt has been made to determine which element dominates the constraint on $q$. This might be important for a detailed study of the elemental abundance influence from nuclear physics inputs, namely, the uncertainty of the reaction cross sections. A study along these lines might be carried in a similar fashion as described in Ref. \cite{NB00}. Weights on the reliability of observational data should also be considered for a more detailed analysis. For example,  constraints arising from $^3$He/H abundance may not be considered trustworthy because of uncertain galactic chemical evolution. On the other hand,  a constraint from the observation of $^3$He/D  is more robust, and so on. Based on our discussion in Section II, it is more important to determine how the non-extensive statistics can modify more stringent conditions during the big bang, such as the modification of weak-decay rates and its influence on the n/p ratio which strongly affects the $^4$He abundance.

\section{Conclusions}
In table \ref{tab1} we present results for the predictions of the BBN  with Maxwellian and non-Maxwellian distributions. The predictions are compared with data from observations reported in the literature. It is evident that the results obtained with the non-extensive statistics strongly disagree with the data. The overabundance of $^7$Li compared to observation gets worse if $q>1$. The three light elements D, $^4$He and $^7$Li constrain the primordial abundances rather well. For all these abundances, a non-extensive statistics with $q>1$ leads to a greater discrepancy with the experimental data.

Except for the case of $^3$He the use of non-extensive statistics with $q<0.5$ does not rule out its validity when the non-Maxwellian BBN results are compared to observations. $^3$He is at present only accessible in our Galaxy's interstellar medium. This means that it cannot be measured at low metallicity, a requirement to make a fair comparison to the primordial generation of light elements. This also means that the primordial $^3$He abundance cannot be determined reliably.  The result presented for the $^3$He abundance  in table \ref{tab1} is quoted from Ref. \cite{Ban02}.  Notice that our analysis does not include the changes that the non-extensive statistics would bring to the n/p conversion rates. The electron distributions would also be expected to change accordingly. This would change the freeze out temperature and a corresponding influence  on the $^4$He abundance.

We conclude that  it does not seem possible to change the Maxwell-Boltzmann statistics to reproduce the observed abundance of light elements in the universe without destroying many other successful predictions of big bang nucleosynthesis. A chi-square fit  of our calculations with the observations of elemental abundance concludes that the non-extensive parameter is constrained to  $q=1^{+0.05}_{-0.12}$. This means that, should a non-Maxwellian distribution due to the use of the Tsallis non-extensive statistics be confirmed (with a sizable deviation from $q=1$), our understanding of the cosmic evolution of the universe would have to be significantly changed.   

\bigskip

We would like to acknowledge beneficial discussions with Alain Coc and Richard Cyburt. This work was partially supported by the US-DOE grants DE-FG02-08ER41533 and DE-FG02-10ER41706, and by the Brazilian agencies, CNPq and FAPESP. One of the authors (C.B.) acknowledges the Helmholtz International Center for FAIR (HIC for FAIR)  for supporting his visit to the GSI Helmholtzzentrum f\"ur Schwerionenforschung, where much of this work was done.

\end{document}